\documentclass[10pt,letterpaper,twocolumn]{article} 

\usepackage{ol2}
\usepackage[draft,implicit=false]{hyperref}
\usepackage{amsmath}

\begin{document}

\twocolumn[ 

\title{Efficient sorting of quantum-optical wave packets by temporal-mode interferometry}

\author{D. V. Reddy$^1$, M. G. Raymer$^{1,*}$, C. J. McKinstrie$^2$}

\address{
$^1$Oregon Center for Optics and Department of Physics, University of Oregon, Eugene, Oregon 97403, USA\\
$^2$Bell Laboratories, Alcatel-Lucent, Holmdel, New Jersey 07733, USA\\
$^*$Corresponding author: raymer@uoregon.edu
}

\begin{abstract}
  Long-distance quantum communication relies on storing and retrieving
  photonic qubits in orthogonal field modes. The available degrees of
  freedom for photons are polarization, spatial-mode profile, and
  temporal/spectral profile. To date, methods exist for decomposing,
  manipulating, and analyzing photons into orthogonal polarization
  modes and spatial modes. Here we propose and theoretically verify
  the first highly efficient method to carry out analogous operations
  for temporally and spectrally overlapping, but field-orthogonal,
  temporal modes. The method relies on cascaded nonlinear-optical
  quantum frequency conversion.
\end{abstract}

 ] 

\noindent Photons -- elementary excitations of the electromagnetic
field -- have four degrees of freedom: helicity and three components
of momentum (which determines frequency through a dispersion
relation). In a beam-like geometry these may be stated as
polarization, spatial-mode profile, and frequency. In order to fully
utilize light for transmitting information in an optical network,
either classical or quantum, it is necessary to be able to sort a beam
of light according to the states of photons associated with each
degree of freedom. Sorting states of light by polarization, center
frequency (wavelength), or spatial-mode profile (transverse parity
\cite{sas03,abou07} or orbital angular momentum
\cite{bar10,pad10,wil13}) are now commonplace. In quantum networks,
other single-photon encoding implementations include the `dual-rail
qubit' \cite{chuang95} and the `time-bin qubit' \cite{marc02}.

A longstanding unsolved problem has been the sorting of light
according to arbitrary states of frequency. By Fourier transformation,
such states can be represented in the time domain, thereby introducing
the idea of `temporal modes' (TMs) -- a set of orthogonal functions of
time that form a complete basis for representing an arbitrary state in
the frequency degree of freedom \cite{smi07}. TM multiplexing does not
offer increased theoretical bounds on channel capacity when compared
to WDM. However, it does offer performance increases for quantum
networks. Efficient sorting and manipulation of temporal modes would
enable concurrent usage of communication pathways by variously encoded
photonic qubits in different `TM-channels', as well as make available
a new basis for encoding quantum information in light -- {\it i.e.}
the `temporal-mode qubit.' Since the dimensionality of the TM-state
space is in principle infinite, single photons could encode quantum
information as multidimensional quantum bits, or qudits, which would
enable increased channel capacity by multimode-entanglement enabled
teleportation \cite{chr12} and device-independent quantum key
distribution based on loophole-free Bell inequality violations
\cite{ver10}. In addition, optical TMs at the same central frequency,
spatial mode and polarization co-propagate over long distances and
remain orthogonal even in the presence of dispersion.

\begin{figure}[thb]
\centering
\includegraphics[width=\linewidth]{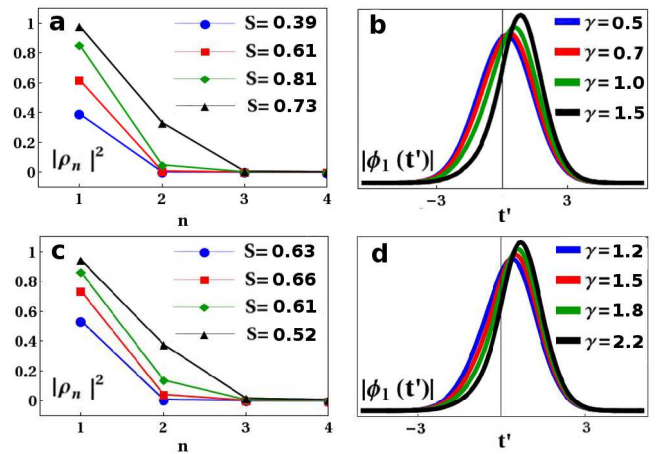}
\caption{\label{fig01}Single-stage TWM (a,b) and FWM (c,d) dominant
  Schmidt-mode conversion efficiencies $|\rho_n|^2$ (a,c) with
  selectivities $S$ listed in the legend. (b,d) {\it s}-channel input
  Schmidt modes. Pump-{\it p} was Gaussian in shape.}
\end{figure}

Here we propose and theoretically verify the first scheme that can
sort TMs with near-$100$\% efficiency, making TM sorting a viable tool
for optical communication. We use an approach based on the observation
that quantum frequency conversion (QFC) is TM-mode selective, as first
pointed out by Eckstein et al. \cite{eck11} and McGuinness et
al. \cite{mcg10a}. In QFC, the signal field of interest interacts with
one (or two) strong laser pulse(s) in a second (or third)-order
nonlinear-optical medium, such as a crystalline waveguide or a silica
optical fiber \cite{ptd12}. The shapes of the laser pulses determine
which TM in the signal will be selected. QFC in materials possessing
$\chi^{(2)}$ or $\chi^{(3)}$ nonlinearity is accomplished using
three-wave mixing (TWM) \cite{huang92,vand04,albota04,rous04} or
four-wave mixing (FWM) Bragg scattering
\cite{mcg10a,mej12b,col12,cla13} respectively. Earlier proposed
implementations of QFC by TWM \cite{eck11} used a single stage of QFC,
which cannot exceed a natural limit for TM `selectivity' of about
$0.8$ \cite{eck11,chr13,red13}. The selectivity of a third-order
nonlinear-optical single-stage version driven by two pump laser pulses
is limited by nonlinear phase-modulation to about $0.65$ (see
Fig. \ref{fig01}). Our proposed scheme, which we call temporal-mode
interferometry (TMI), uses cascaded stages of QFC, and can provide
close to $100$\% TM selectivity, opening the door for manipulation of
TM qubits or multi-level qudits. The method may also be used for TM
channel multiplexing in classical optical telecom.

To model the QFC process, we designate the participating frequency
channels by the letters {\it s}, {\it r} for the signal photons and
{\it p} (and {\it q}) for the strong pump field(s). We denote
normalized field envelopes in the {\it j}-channel by $A_j(z,t)$ and
the group slowness (inverse group velocity) by
$\beta'_j\equiv[d\beta/d\omega]|_{\omega=\omega_j}$, and we neglect
higher-order dispersion, which is valid for narrow-band pulses. Then
the equations of motion for QFC are \cite{myers95,mej12b}:

\begin{subequations}
\label{eomfwm}
\begin{align}
(\partial_z+\beta'_p\partial_t)A_p&=i(\gamma/2)\delta_F\left[\left|A_p\right|^2+2\left|A_q\right|^2\right]A_p,\\
(\partial_z+\beta'_q\partial_t)A_q&=i(\gamma/2)\delta_F\left[2\left|A_p\right|^2+\left|A_q\right|^2\right]A_q,\\
(\partial_z+\beta'_r\partial_t)A_r&=i\gamma A_pA_q^*A_s\notag\\
&+i\gamma\delta_F\left[\left|A_p\right|^2+\left|A_q\right|^2\right]A_r,\\
(\partial_z+\beta'_s\partial_t)A_s&=i\gamma A_p^*A_qA_r\notag\\
&+i\gamma\delta_F\left[\left|A_p\right|^2+\left|A_q\right|^2\right]A_s,
\end{align}
\end{subequations}

\noindent where the self- and cross-phase modulation factor
$\delta_F$, is $0$ for TWM and $1$ for FWM. $\gamma$ is the product of
the effective $\chi^{(2)}$ or $\chi^{(3)}$ nonlinearity and the square
roots of the pump-pulse energies. For TWM, $A_q(z,t)\equiv 1$. The
solution for Eq. (\ref{eomfwm}) can be written in terms of Green
functions $G_{ij}(t,t')$. For $j\in \{r,s\}$ and medium length $L$:

\begin{equation}
A_j(L,t)=\int^\infty_{-\infty}\mathrm{d}t'\sum\limits_{k=r,s}G_{jk}(t,t')A_k(0,t').\label{eq07}
\end{equation}

The four kernels $G_{ij}(t,t')$ comprise a generalized
beam-splitter transformation, representing frequency conversion or
non-conversion of each of the two possible input fields of distinct
carrier frequencies \cite{ray10}. 

For each $G_{ij}(t,t')$, we can perform numerically a singular-value
decomposition (SVD), also called a Schmidt decomposition, into a set
of singular-value functions (temporal Schmidt modes) with associated
singular values (Schmidt coefficients), which we denote by $\rho_n$
and $\tau_n$, which satisfy $|\rho_n|^2+|\tau_n|^2=1$. $|\rho_n|^2$ is
the frequency conversion efficiency (CE) of the $n^\mathrm{th}$ mode,
and can be interpreted as the probability of QFC in the case of
single-photons. The modes are numbered in decreasing order of the
Schmidt coefficients. We denote by $\phi_n(t)$ the {\it s}-input
modes, and $\Psi_n(t)$ are {\it r}-output modes. In addition, there
are {\it r}-input modes $\psi_n(t)$ and {\it s}-output modes
$\Phi_n(t)$. All the Green functions can be expressed in terms of
these four mode sets as follows \cite{ray10}:

\begin{subequations}
\begin{align}
G_{rr}(t,t')&=\sum\limits^\infty_{n=1}\tau_n\Psi_n(t)\psi_n^*(t'),\\
G_{rs}(t,t')&=\sum\limits^\infty_{n=1}\rho_n\Psi_n(t)\phi_n^*(t'),\\
G_{sr}(t,t')&=-\sum\limits^\infty_{n=1}\rho_n\Phi_n(t)\psi_n^*(t'),\\
G_{ss}(t,t')&=\sum\limits^\infty_{n=1}\tau_n\Phi_n(t)\phi_n^*(t').
\end{align}
\end{subequations}

The chief objective is to design a QFC device that can selectively
frequency convert or `drop' the first Schmidt mode with unit
efficiency, whilst allowing $100$\% unconverted transmission of all
orthogonal modes. The principal figure of merit for such a drop device
is the `selectivity,' \cite{red13} defined as $S =
|\rho_1|^4/\sum^\infty_{j=1}|\rho_j|^2$.

\begin{figure*}[t]
\centering
\includegraphics[width=0.8\linewidth]{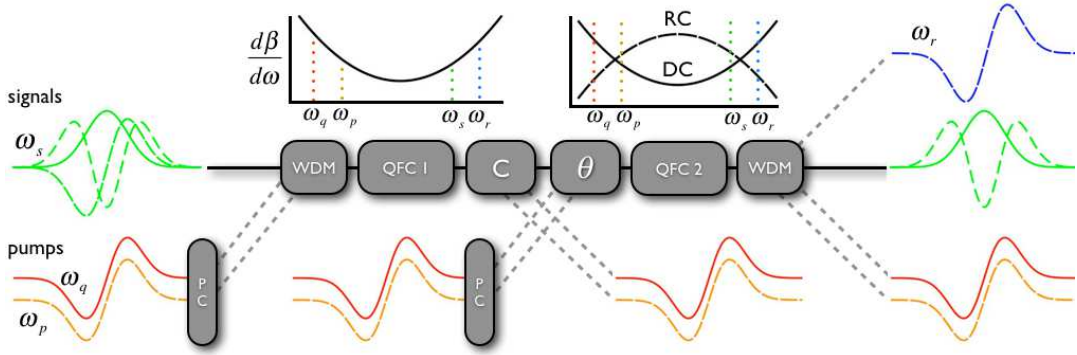}
\caption{\label{fig02}Temporal-mode interferometer using two nonlinear
  media (QFC 1 and QFC 2) with identical (DC) or opposite-sign (RC)
  dispersion. By choosing the pump-pulse shapes, pump powers and phase
  shift $\theta$, one can selectively frequency convert a specific
  {\it s} (green) TM into an {\it r} (blue) TM at a different central
  frequency, whilst not affecting temporally-orthogonal {\it s}-input
  TMs. The {\it q}-channel is used only for $\chi^{(3)}$
  implementations. WDM stands for wavelength-division multiplexer. PC
  stands for pre-chirp module, which are necessary for $\chi^{(3)}$
  implementations. The coupler C contains frequency dependent delays
  for the DC case.}
\end{figure*}

The best previous attempts at optimizing selectivity using a
single-stage QFC scheme have converged on frequency choices that are
group-velocity matched, namely
$\beta'_p=\beta'_s\neq\beta'_r$($=\beta'_q$)\cite{mej12b,col12,red13}.
Eq. (\ref{eomfwm}) been fully solved analytically for this case for
TWM \cite{red13}. In addition, for optimum results in FWM, the medium
is required to be long enough for a complete inter-pump-pulse
collision (no-overlap to no-overlap)\cite{mej12b}. However, both TWM
and FWM schemes have encountered limits that are universal to generic
systems governed by coupled-mode equations such as Eq. (\ref{eomfwm}),
and have yielded selectivities limited to around 0.8 and Schmidt-mode
profiles that are severely temporally skewed relative to the
group-velocity-matched pump shapes (Fig. \ref{fig01}). The limit to
single-stage selectivity and the Schmidt-mode temporal-skewness occur
due to oscillatory distortion of the Green functions at higher
$\gamma$ akin to Burnham-Chiao ringing \cite{Burnham1969,red13}. These
distortions are minimal at lower CE (Fig. \ref{fig01}).

We have discovered that using a two-stage interferometric scheme
overcomes both of these limitations. Each stage is configured for
$50$\% CE for the first Schmidt mode (Fig. \ref{fig02}), and functions
as a $50/50$ beam splitter with the {\it r}- and {\it s}-frequency
channels representing its two input and output arms. An {\it s}-input
photon in the first Schmidt mode in the first stage will be partially
coherently frequency converted into the {\it r}-channel with a phase
picked up from the pump field(s). If all fields are allowed to
participate in QFC in the second stage at the right relative phases,
the two effective beam splitters will function as a frequency-shifting
Mach-Zehnder interferometer, allowing for complete FC or complete
back-coversion of the photon. The effect operates only on the first
Schmidt mode, as the higher-order modes have negligible CE in both
stages. We call the scheme `temporal-mode interferometry.' There are
two configurations for which the four pulsed fields interact in both
stages: 1) `reversed collision' (RC), in which the second-stage
dispersion is inverted relative to the first, {\it i.e.}
$\beta'_r|^{(1)}=\beta'_s|^{(2)}$ and vice versa, such that the
relative speeds of the pulses are reversed, and 2) `double collision'
(DC), in which the second-stage dispersion is identical to the first,
but the fast pulses are time delayed relative to the slow pulses in
between the stages so that they walk through each other again.

For the two-stage TMI, the combined Green function
kernel $G_{rs}(t,t')$ is given by the interferometric equation:

\begin{equation}
\begin{split}
G_{rs}(t,t')=\int_{-\infty}^\infty\mathrm{d}t''\big[G^{(2)}_{rs}(t,t'')G^{(1)}_{ss}(t'',t')\\
+e^{i\theta}G^{(2)}_{rr}(t,t'')G^{(1)}_{rs}(t'',t')\big],
\end{split}\label{eq15}
\end{equation}

\begin{figure*}[t]
\centering
\includegraphics[width=\linewidth]{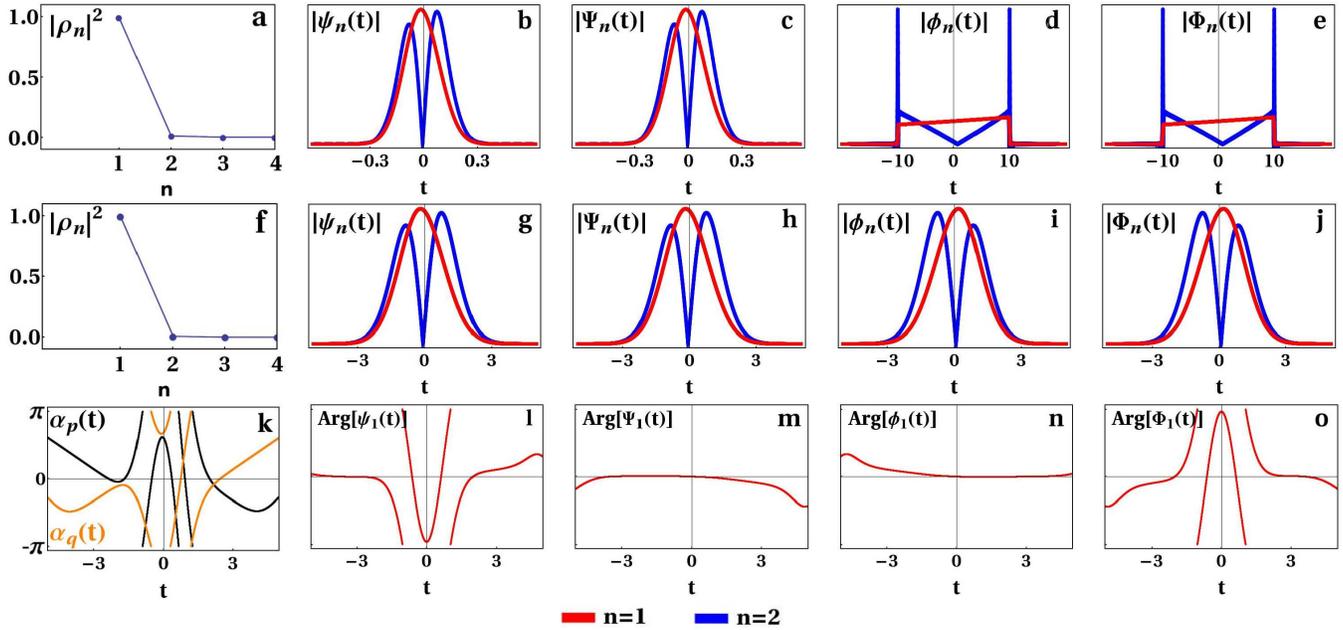}
\caption{\label{fig04}Conversion efficiencies and Schmidt modes for
  (a,b,c,d,e) TWM-TMI, (f,g,h,i,j) FWM-TMI, and (l,m,n,o) first
  Schmidt-mode phase profiles for FWM-TMI, all with Gaussian
  pumps. Plots (a,f) show the conversion efficiencies of the first
  four Schmidt modes. The first two Schmidt modes for the
  corresponding processes are shown for {\it r}-input (b,g), {\it
    r}-output (c,h), {\it s}-input (d,i), and {\it s}-output
  (e,j). (k) shows the first-stage pump pre-chirps used for FWM-TMI.}
\end{figure*}

\noindent where the parenthetical superscripts indicate the stage. A
control phase $\theta$ is externally applied to any one of the
channels in between stages. Equation (\ref{eq15}) illustrates the need
for inter-stage temporal mode matching between the output Schmidt
modes of the first stage and the input Schmidt modes of the second
stage for the scheme to work. For TWM-TMI, this is best achieved by
the RC configuration. Figure \ref{fig04}(a)-\ref{fig04}(e) presents
the CE and Schmidt modes for the RC configuration TWM-TMI with
$\zeta=|\beta'_r-\beta'_s|L/\tau_p = 200$, where $L$ is the per stage
medium length and $\tau_p$ is the pump width. The selectivity was
computed to be $0.9846$ ($|\rho_1|^2=0.9975$,
$|\rho_2|^2=0.0110$). $\zeta$ is also the ratio of time-widths of {\it
  r}- and {\it s}-Schmidt modes
(Fig. \ref{fig04}(b)-\ref{fig04}(e)). The selectivity of the RC-TWM
configuration improves asymptotically with $\zeta$. A $\zeta$ of $200$
can be realized, for example, in a typical $5$-cm long PPLN waveguide
with a $70$-fs pump-pulse and signal wavelengths at $795$ nm and
$1324$ nm.

The first-stage output pulses have the correct relative time delays
for the second-stage input in the RC configuration. However, TWM-TMI
can also be implemented in the DC configuration, using appropriate
frequency dependent delay lines in between stages. For $\zeta=200$, it
yielded a slightly lower selectivity of $0.9805$ ($|\rho_1|^2=0.9957$,
$|\rho_2|^2=0.0134$) due to relatively inferior inter-stage
mode-matching. TWM-TMI can be easily extended to an arbitrary number
of stages with arbitrary ordering of RC/DC stage-interfaces and
appropriately lower pump powers. Such schemes will be explored in a
later publication.

Nonlinear phase-modulation in FWM severely affects the Schmidt-mode
phase-profiles, and restricts FWM-TMI to the RC configuration. The
pumps must be pre-chirped with specific phase-profiles \cite{mej12b}
for each stage to enhance inter-stage mode-matching. Figure
\ref{fig04}(f)-\ref{fig04}(o) shows the numerically computed CE,
Schmidt-mode amplitudes, and phases corresponding to FWM-TMI with
selectivity $0.9873$ ($|\rho_1|^2=0.9973$, $|\rho_2|^2=0.0082$). The
pump pre-chirps (Fig. \ref{fig04}(k)) were specifically chosen to
yield flat phase profiles for the {\it r}-output (Fig. \ref{fig04}(m))
and {\it s}-input (Fig. \ref{fig04}(n)) Schmidt modes. The results
were computed for medium lengths that were long enough for complete
pump-pulse collisions in each stage
($|\beta'_r-\beta'_s|L/(\tau_p+\tau_q)=5$), a condition easily
satisfied by use of highly nonlinear fibers for QFC
\cite{mcg10a}. FWM-TMI, besides yielding good selectivity, gives the
ability to choose the {\it s}- and {\it r}-channel Schmidt-mode shapes
independently of each other (by reshaping both pumps), allowing for
efficient mode-selective single-photon temporal reshaping. FWM-TMI, in
contrast to TWM-TMI, does not enforce a high temporal-width ratio
between the signal channels. However, nonlinear phase-modulation
hinders the extension of FWM-TMI into multi-stage schemes.

In conclusion, TMI can selectively manipulate photonic temporal mode
components of any shape accessible by our ability to reshape strong
pump pulses. Modular TMI devices also possess a phase control
($\theta$) that can suppress QFC completely and make the device
transparent on demand. The addition of temporal-mode analysis and
multiplexing to polarization and OAM analysis and multiplexing
completes the toolkit for encoding information on quantum and
classical states of light.

We thank L. Mejling, K. Rottwitt and J. Nunn for helpful
discussions. This work was supported by the National Science
Foundation through EPMD and GOALI, grant ECCS-1101811.

\end{document}